\documentclass[12pt]{article}
\usepackage{array,multicol}
\usepackage{afterpage,float,flafter}
\usepackage[dvips]{epsfig,rotating}
\usepackage{pifont,fancybox}
\usepackage{latexsym}

\newcommand{\be}{\begin{equation}}
\newcommand{\ee}{\end{equation}}

\newcommand{\bea}{\begin{eqnarray}}
\newcommand{\eea}{\end{eqnarray}}
\newcommand{\eq}[1]{eq.~(\ref{#1})}

\newcommand{\gsim}{\ \rlap{\raise 2pt\hbox{$>$}}{\lower 2pt \hbox{$\sim$}}\ }
\newcommand{\lsim}{\ \rlap{\raise 2pt\hbox{$<$}}{\lower 2pt \hbox{$\sim$}}\ }

\newcommand{\matr}{\left( \begin{array}}
\newcommand{\ematr}{\end{array} \right)}

\newcommand{\anfis}[1]{Ann. Phys. #1}

\def\bq{\begin{quote}}
\def\eq{\end{quote}}
\def\ben{\begin{enumerate}}
\def\een{\end{enumerate}}

\def\p{{\bf p}}

\def\x{{\bf x}}
\def\y{{\bf y}}
\def\xp{{\bf x^{\prime}}}

\def\ie{{\it i.e.}}

\def\e{{\rm e}}

\def\H{{\cal H}}

\def\p{{\bf p}}
\def\q{{\bf q}}
\def\x{{\bf x}}
\def\y{{\bf y}}

\def\Dslash{\not{\hbox{\kern-4pt $D$}}}
\def\dslash{\kern-4pt \not{\hbox{\kern-2pt $\partial$}}}
\def\pslash{\not{\hbox{\kern-2pt p}}}

\makeatletter
\@addtoreset{equation}{section}
\makeatother

\title{
\vspace*{-2.0cm}
\begin{flushright}
\normalsize{
FERMILAB-Pub-02/266-T
}
\end{flushright}
\vspace*{1.0cm}
{A Model of $CPT$ Violation for Neutrinos}
\vspace*{0.8cm}
\author{\large\textbf
{Gabriela~Barenboim$^a$ and Joseph~Lykken$^{a,b}$}\\ 
\\
$^a$\normalsize\emph{Fermi National Accelerator Laboratory,
P.O. Box 500, Batavia, IL 60510, USA }\\
$^b$\normalsize\emph{Enrico Fermi Institute and
Dept. of Physics,}\\ 
\normalsize\emph{University of Chicago,
5640 S. Ellis Ave., Chicago, IL 60637, USA}\\ }
}

\begin{document}
\maketitle

\vspace*{2cm}

\begin{abstract}
Any local relativistic quantum field theory of Dirac-Weyl fermions
conserves $CPT$. Here we examine whether a simple nonlocal
field theory can violate $CPT$. We construct a new relativistic field
theory of fermions, which we call
``homeotic'', which is nonlocal but causal and Lorentz invariant.
The free homeotic theory is in fact equivalent to
free Dirac theory. We show that a homeotic theory with
a suitable nonlocal four-fermion interaction is causal
and as a result has a well-defined perturbative S-matrix.
By coupling a right-handed homeotic fermion to
a left-handed Dirac-Weyl fermion, we obtain a causal
theory of $CPT$-violating neutrino oscillations.
\end{abstract}

\thispagestyle{empty}
\newpage

\section{Introduction}
$CPT$ violating neutrino masses allow the possibility 
\cite{Murayama}-\cite{Nos2}
of reconciling the LSND \cite{lsnd}, atmospheric \cite{atm_sk}, 
and solar oscillation \cite{solar, SNO_solar}
data without resorting to sterile neutrinos. As argued in
our previous work \cite{Nos1}, 
there are good reasons to imagine that $CPT$ violating
dynamics couples directly to the neutrino sector, but not
to other Standard Model degrees of freedom. For $CPT$ violating
Dirac mass splittings $\Delta m$ on the order of 1 eV or less, the feed-through
to other Standard Model processes is completely negligible,
suppressed by $G_F(\Delta m)^2$ times loop factors.

$CPT$ violation in the neutrino sector would be a very
exciting development, and raises important theoretical issues.
$CPT$ is conserved quite generally in local relativistic quantum
field theory \cite{Steater}, and even in large classes of nonlocal effective
field theories \cite{Efimov}. While it is straightforward to construct field
theories which break $CPT$ via spontaneous violation of Lorentz
invariance \cite{Kostelecky}, 
this would seem to entail other drastic consequences \cite{Mocioiu},
unless restricted to the context of cosmology. As far as is known,
$CPT$ is not gauged in string theory, and thus there is no
general argument that it should be respected at high energies
\cite{stringycpt}.
However this begs the question of how $CPT$ violation
will appear in the effective low energy theory.

If $CPT$ violation is observed in
the neutrino sector, must we give up either on-shell Lorentz
invariance or effective quantum field theory in our description of
the low energy world? To answer this question, we have constructed
what we believe is the minimal relativistic quantum field theory model of
explicitly $CPT$ violating neutrinos. The $CPT$ violation arises
from a Dirac type local bilinear coupling
$\bar{\nu}_LN_R + \bar{N}_R\nu_L$, where $\nu_L(x)$ is the
usual left-handed Dirac-Weyl neutrino that occupies an
electroweak doublet with a Standard Model charged lepton.
The other field, $N_R(x)$, is a Standard Model singlet
fermion with novel properties, which we call {\it homeotic}.

In the next section we develop the relativistic quantum
field theory of homeotic fermions, in analogy to that of
Dirac and Dirac-Weyl fermions. Homeotic fermions obey
standard Fermi statistics, and we exhibit interacting
lagrangian field theories for homeotic fermions which are
both unitary and causal. Not surprisingly, the theory of
free homeotic fermions is physically equivalent to the usual
field theory of free Dirac fermions: it has the same degrees of
freedom, the same hamiltonian and
the same symmetries. However the
homeotic free field theory has a nonlocal lagrangian, and the
homeotic fields have noncanonical anticommutation relations.

By adding nonlocal relativistic four-fermion interactions to
the free homeotic theory, we obtain a causal interacting field
theory with a well-defined unitary S-matrix. Indeed this interacting
homeotic theory appears to have the same S-matrix as Dirac theory with a
local four-fermion interaction. 
The homeotic theory with only local four-fermion interactions,
on the other hand,
is acausal and does not have a well-defined S-matrix.
Naively this theory would violate crossing symmetry at tree level.

The homeotic field theories
are $CPT$ invariant, but $CPT$ is of necessity realized by a
different symmetry operator than for Dirac theory. The
bilinear mass term discussed above, which couples a
left-handed Dirac-Weyl fermion to a right-handed homeotic
fermion, thus violates $CPT$.

The resulting field theory provides the
simplest relativistic quantum field theory model
for $CPT$ violating neutrino oscillations. There are
still issues in understanding the off-shell
description of neutrino propagation, as discussed in the
work of Greenberg \cite{Greenberg}. But there is great virtue in having
a simple explicit model.

\section{Free homeotic fermions}

In this section we develop the field theory of free homeotic
fermions, in analogy with free Dirac theory. We employ the notation
and conventions of Peskin and Schroeder \cite{Peskin}. 
Recall that in Dirac theory
we introduce positive and negative frequency solutions of the Dirac
equation:
\bea
\psi_+(x) &= u_+(p)\e^{-ip\cdot x} \, ,\quad p^2=m^2,\quad p_0 > 0; \cr
\psi_-(x) &= u_-(p)\e^{-ip\cdot x} \, ,\quad p^2=m^2,\quad p_0 < 0,
\eea
and the 4-component spinors $u_+(p)$, $u_-(p)$ satisfy
\be
(\pslash - m)u_{\pm}(p) = 0.
\ee
Note that $u_+(p)$, $u_-(p)$ are usually written
as $u(p)$, $v(-p)$, respectively, where $u(p)$, $v(p)$ obey the
equations
\bea
(\pslash - m)u(p) = 0,\cr
(\hbox{--$\pslash$} - m)v(p) = 0.
\eea
The homeotic theory is built from spinors $u_+(p)$, $u_-(p)$ which
satisfy
\be
(\pslash - m\epsilon(p_0))u_{\pm}(p) = 0,
\label{heom}
\ee
where $\epsilon(p_0)$ is the sign function of the standard
delta calculus. Note that solutions of Eq.~(\ref{heom})
automatically satisfy the Klein-Gordon equation and thus
the usual on-shell dispersion relation
$p^2 = m^2$. This in turn implies that the equation of motion
Eq.~(\ref{heom}) is Lorentz covariant, since
$\epsilon(p_0)$ is Lorentz invariant for timelike 4-momenta.

Rewriting $u_+(p)$, $u_-(p)$ as $u(p)$, $\tilde{u}(-p)$,
respectively, we see that homeotic fermions are built out
two sets of $u$-spinors, both satisfying the usual positive
frequency relation
\be
(\pslash - m)u(p) = (\pslash - m)\tilde{u}(p) = 0.
\ee

On-shell fields of the homeotic theory are assembled from
the obvious plane wave expansions:
\bea
\psi(x) &= \displaystyle{\int {d^3p\over (2\pi)^3}
{1\over\sqrt{2E_{\p}}} \sum_s} \left[
\displaystyle{a_{\p}^s u^s(p) \e^{-ip\cdot x} +
b_{\p}^{s\dagger} \tilde{u}^s(p) \e^{ip\cdot x}} \right],
\cr\cr
\psi^{\dagger}(x) &= \displaystyle{\int {d^3p\over (2\pi)^3}
{1\over\sqrt{2E_{\p}}} \sum_s} \left[
\displaystyle{b_{\p}^s \tilde{u}^{s\dagger}(p) \e^{-ip\cdot x} +
a_{\p}^{s\dagger} u^{s\dagger}(p) \e^{ip\cdot x}} \right],
\eea

where $\p$ denotes 3-momenta, and $s = 1$, 2, is a spin label.
The homeotic field $\psi(x)$ is a solution of the equation
of motion

\be
i\dslash \psi(t,\x) = -{im\over\pi}\,{\bf P}\int dt^{\prime}\,
{1\over t - t^{\prime}} \,\psi(t^{\prime},\x) \;,
\ee
where {\bf P} denotes the principal value integral, which
we will now assume throughout.  

This equation of motion can be obtained from variation of the
following nonlocal action:
\be
{\bf S} = \displaystyle{\int d^4x\, \bar{\psi}(x)i\dslash \psi(x)
+ {im\over\pi}\int d^3x\int dtdt^{\prime}\,
\bar{\psi}(t,\x)\,{1\over t-t^{\prime}}\,\psi(t^{\prime},\x) .}
\label{haction}
\ee
This action is Lorentz invariant, as can be seen by rewriting the
mass term as
\be
{im\over\pi}\int d^3xd^3x^{\prime}\int dtdt^{\prime}\,
\bar{\psi}(t,\x)\,{\delta^{(3)}(\x - \xp)
\over t-t^{\prime}}\,\psi(t^{\prime},\xp) ,
\label{hactionb}
\ee
and observing that the integral
\be
\int d^3xdt\, {\delta^{(3)} (\x)\over t} f(t,\x)
\label{boostinv}
\ee
with an arbitrary scalar function $f(t,\x)$, is invariant under boosts.

To quantize this theory, we treat $a_{\p}^s$ and $b_{\p}^s$ as
anticommuting Fock operators obeying
\be
\displaystyle{
\{a_{\p}^r,a_{\q}^{s\dagger}\} =
\{b_{\p}^r,b_{\q}^{s\dagger}\} =
(2\pi )^3\delta^{(3)}(\p - \q)\delta^{rs} \,.}
\label{fermistat}
\ee
These relations guarantee conventional Fermi statistics.
The Hilbert space built from the corresponding Fock vacuum supports
a standard representation of the Lorentz group.

The homeotic fields $\psi(x)$ do not have canonical anticommutation
relations:
\bea
\left\{ \psi_a^{\dagger}(x), \psi_b(y) \right\}_{x^0=y^0}&=
&\displaystyle{\int {d^3p\over (2\pi )^3} \sum_s
{1\over 2E_{\p}} \left[
u_a^{s\dagger}(p)u_b^s(p)\e^{ip\cdot (x-y)}
+ \tilde{u}_a^{s\dagger}(p)\tilde{u}_b^s(p)
\e^{-ip\cdot (x-y)} \right]_{x^0=y^0}} \cr
&&\cr
&= &\delta^{(3)}(\x - \y )\delta_{ab}
+ m(\gamma^0)_{ab}\left[ D(x-y) + D(y-x) \right]\;,
\label{hcom}
\eea
where $D(x-y)$ is the familiar function invariant under proper
orthochronous Lorentz transformations:
\be
D(x-y) = \int {d^3p\over (2\pi )^3} {1\over 2E_{\p}} \e^{-ip\cdot (x-y)} \; .
\ee

The general anticommutator function \cite{Bogoliubov} 
of the homeotic theory is given by
\be
\displaystyle{\left\{ \bar{\psi}_a(x), \psi_b(y) \right\} =
\left(i\dslash_x + m\right)_{ab}D(x-y) +
\left(i\dslash_y + m\right)_{ab}D(y-x) \; ,}
\label{genhcom}
\ee
which differs from the Dirac result by the sign of the first term
proportional to $m$.

An important feature of Eq.~(\ref{hcom}) is that the noncanonical
term occurs only for anticommuting homeotic fermion components of opposite
chiralities:
\bea
\left\{ \psi_{La}^{\dagger}(x), \psi_{Lb}(y) \right\}_{x^0=y^0}&=
&\delta^{(3)}(\x - \y )\delta_{ab}\;, \cr
\left\{ \psi_{Ra}^{\dagger}(x), \psi_{Rb}(y) \right\}_{x^0=y^0}&= 
&\delta^{(3)}(\x - \y )\delta_{ab}\;, \cr
\left\{ \psi_{La}^{\dagger}(x), \psi_{Rb}(y) \right\}_{x^0=y^0}&= 
&m\delta_{ab}\left[ D(x-y) + D(y-x) \right]\;.
\label{hcomchiral}
\eea
We stress again that homeotic fermions obey
conventional Fermi statistics; this was guaranteed from the start
by Eqs.~(\ref{fermistat}).

Note that in the strict nonrelativistic limit
($\vert \p \vert \rightarrow 0$ with $m$ fixed) the
equal-time anticommutators reduce to
\bea
\left\{ \psi_{La}^{\dagger}(x), \psi_{Lb}(y) \right\}_{x^0=y^0}&=
&\delta^{(3)}(\x - \y )\delta_{ab}\;, \cr
\left\{ \psi_{Ra}^{\dagger}(x), \psi_{Rb}(y) \right\}_{x^0=y^0}&= 
&\delta^{(3)}(\x - \y )\delta_{ab}\;, \cr
\left\{ \psi_{La}^{\dagger}(x), \psi_{Rb}(y) \right\}_{x^0=y^0}&= 
&\delta^{(3)}(\x - \y )\delta_{ab}\;,
\label{nrhcomchiral}
\eea
where the third anticommutator would vanish if these were Dirac fermions.
Evidently one of the zero momentum modes, $\psi_L$$+$$\psi_R$,
is canonically quantized, while the other one,
$\psi_L$$-$$\psi_R$, has a vanishing anticommutator.
Thus zero mode quantization requires special care in homeotic theory.

The symmetrized stress-energy tensor derived from the action
(\ref{haction}) has the same form as in Dirac theory and is
conserved on-shell:

\be
\Theta^{\mu\nu}(x) = \displaystyle{{i\over 4}
\left[ \bar{\psi}\gamma^{\mu}\partial^{\nu}\psi
- \partial^{\nu}\bar{\psi}\gamma^{\mu}\psi
+ (\mu \leftrightarrow \nu )\right].}
\ee

From this expression we obtain the hamiltonian of the free homeotic
theory:
\bea
{\bf H} &= \int d^3x\, \H (x)  = \int d^3x\, \Theta^{00}(x)
= \int d^3x\, {i\over 2}\left[ \psi^{\dagger}\dot{\psi}
-\dot{\psi}^{\dagger}\psi \right]\cr
&= \displaystyle{\int {d^3p\over (2\pi)^3} \sum_s
E_{\p} \left[ a_{\p}^{s\dagger}a_{\p}^s + b_{\p}^{s\dagger}b_{\p}^s
\right]\;, }
\label{hham}
\eea
where we have dropped the normal ordering constant.
Similarly, we obtain the momentum operator:
\be
{\bf P}^{\bf i} = \int d^3x\, \Theta^{0i}(x)
= \displaystyle{\int {d^3p\over (2\pi)^3} \sum_s
{\p}^{\bf i} \left[ a_{\p}^{s\dagger}a_{\p}^s + b_{\p}^{s\dagger}b_{\p}^s
\right]\;. }
\label{hmom}
\ee
The boost operator is defined by
\be
{\bf K}^{\bf i} = \int d^3x\, \left( x^i\Theta^{00}
- x^0\Theta^{0i} \right) 
= -t{\bf P}^{\bf i} + \int d^3x\, {\bf x}^{\bf i}
\H (x)\; .
\label{hboost}
\ee
As a straightforward but nontrivial check of Lorentz invariance, one
can verify that the definitions (\ref{hham}-\ref{hboost}) together with
the anticommutation relations (\ref{genhcom}) produce the on-shell
relation
\be
\left[ {\bf K}^{\bf i}, {\bf H} \right] = i{\bf P}^{\bf i} \; .
\ee 

The action (\ref{haction}) also has a global $U(1)$ symmetry under
phase rotations of $\psi(x)$. The conserved charge is
\be
\displaystyle{Q = \int {d^3p\over (2\pi)^3} \sum_s
\left[ a_{\p}^{s\dagger}a_{\p}^s - b_{\p}^{s\dagger}b_{\p}^s
\right]\;. }
\label{ourcharge}
\ee
These symmetry operators (\ref{hham}-{\ref{ourcharge}) are
the same as in Dirac theory. The matrix elements of the free homeotic
theory are in one-to-one equivalence with matrix elements of
free Dirac theory.

\section{Interacting homeotic fermions}

In the free homeotic theory, the conserved charge $Q$
of Eq.~(\ref{ourcharge}) is not given by the
spatial integral of a local charge density. Instead (following a
trick of Pauli's \cite{Pauli}) we find the expression
\bea
Q = \int d^3x\, \bar{\psi}\gamma^0\psi ~~~~~~~~~~~~~~~~~~~~~~~~~~~~~~~
~~~~~~~~~~~~~~~~~~~~~~~~~~~~~~~~~~~~~~\cr
+\displaystyle{{m\over\pi}\int^t_{-\infty}dt_1 \int^{\infty}_{-\infty} dt_2
\int d^3x\, {1\over t_1 -t_2}\left[
\bar{\psi}(t_1,\x)\psi(t_2,\x)+\bar{\psi}(t_2,\x)\psi(t_1,\x) \right]}\; .
\label{ourchargeb}
\eea
The corresponding conserved current is:
\be
J^{\mu}(x) = J_D^{\mu}(x)
+\displaystyle{\delta^{\mu 0}
{m\over\pi}\int^t_{-\infty}dt_1 \int^{\infty}_{-\infty} dt_2
\, {1\over t_1 -t_2}\left[
\bar{\psi}(t_1,\x)\psi(t_2,\x)+\bar{\psi}(t_2,\x)\psi(t_1,\x) \right]}\; ,
\label{ourcurrent}
\ee
where $J_D^{\mu}(x) = \bar{\psi}\gamma^{\mu}\psi$ is the conserved
current of Dirac theory.

As with any conserved abelian current in the canonical formalism, the
following current algebra relations hold:
\bea
\left[ J^0(x), J^0(y) \right]_{x^0=y^0} &=&0 \;, \cr
\left[ J^0(x), J^i(y) \right]_{x^0=y^0} &=&0 \;, \cr
\left[ J^i(x), J^j(y) \right]_{x^0=y^0} &\propto&\delta^{(3)}(\x - \y) \;, 
\label{hcurrentalg}
\eea
where it is understood that in the canonical formalism we fail to
pick up the expected Schwinger term in the second
and third expression \cite{Jackiw}.

\subsection{A causal interacting theory}

We can construct a model of interacting homeotic fermions by
introducing a simple current-current four-fermion interaction:
\be
H_I(t) = \int d^3x \H_{I}(x) = \int d^3x J^{\mu}(x)J_{\mu}(x) \; .
\label{ourfourfermint}
\ee

Since the homeotic fields do not obey canonical anticommutation
relations, we must carefully define what we mean by a relativistic
quantum field theory. The obvious approach is to go to the
interaction picture and attempt to construct a perturbative
S-matrix. One can easily verify that the homeotic free hamiltonian $H_0$
generates conventional interaction picture time evolution:
\be
H_I(t) = \e^{iH_0t}\,H_I\,\e^{-iH_0t}\; ,
\ee
where it is understood that the mass parameter which appears
in $H_0$, and thus in the anticommutation relations
(\ref{hcom}), is the physical mass.

The perturbative S-matrix is defined as the matix elements of the
Dyson series:
\be
{\cal S} = 1 + \sum^{\infty}_{n=1} {(-i)^n\over n!}
\int d^4x_1\cdots d^4x_n\; T\left\{
\H_I(x_1)\cdots\H_I(x_n)\right\} \; ,
\label{dysonseries}
\ee
where $T$ denotes time-ordering. The current algbera relations
(\ref{hcurrentalg}) imply the causality condition
\be
\left[ \H_I(x), \H_I(y) \right]_{x^0=y^0} \propto\delta^{(3)}(\x - \y) \; ,
\label{causalitycond}
\ee
which in turn guarantees the Lorentz invariance of the time-ordering
in (\ref{dysonseries}). Together with the properties of the free
homeotic theory detailed above, this is sufficient to prove that
our interacting homeotic theory has a unitary Lorentz invariant 
S-matrix. This in turn can be regarded as proof that the homeotic
theory is a sensible relativistic quantum field theory. 
Indeed this S-matrix appears equivalent to what we would
have obtained from Dirac theory with an analogous current-current
interaction. 

This interacting homeotic theory is quite a {\it rara avis}:
despite the fact that the lagrangian is highly nonlocal the
field theory is unitary, causal, and Lorentz invariant.
It also exhibits crossing symmetry and conserves $CPT$. However,
$CPT$ is not realized in the standard way. 
The $CPT$ operator on Dirac spinors
is $\gamma^5$, but on homeotic spinors it is $\gamma^2 \gamma^5$.
This difference comes from the charge conjugation
operation (time reversal and parity are realized in the
standard way) and is the direct consequence of the lack of
$v(p)$ spinors in the homeotic theory. Whereas in Dirac theory
charge conjugation relates a $u(p)$ spinor to $\gamma^2v^*(p)$,
in homeotic theory it merely exchanges $u(p)\leftrightarrow\tilde{u}(p)$.
Note that the nonstandard $CPT$ properties of homeotic theory are
essential to reconciling $CPT$ conservation with the existence
of the conserved current (\ref{ourcurrent}): both of the terms in
(\ref{ourcurrent}) are even under $CPT$, in contrast to the Dirac
case where $\bar{\psi}\gamma^{\mu}\psi$ is $CPT$ odd.   

The perturbative Feynman rules of this theory are easily
derived. The propagator is defined unambiguously following
the discussion of Weinberg \cite{Weinberg}:
\be
-i\Delta^h_{ab}(x,y) = \theta (x^0-y^0)\left\{\psi^+_a(x),
{\psi^+_b}^{\dagger}(y)\right\}
- \theta (y^0-x^0)\left\{ {\psi^-_b}^{\dagger}(y),
\psi^-_a(x)\right\} \; ,
\ee
where $\psi^{\pm}(x)$ denote the positive/negative frequency
contributions to $\psi(x)$. A simple calculation gives
\be
\Delta^h_{ab}(x,y) = \displaystyle{\int {d^4p\over (2\pi)^4}
\,{i\left[\pslash + m\epsilon(p_0)\right]_{ab}\over p^2 - m^2 +i\epsilon}
\,\e^{-ip\cdot(x-y)} }\; .
\label{hpropagator}
\ee
This propagator is not equivalent to the usual Feynman propagator
and does not give causal propagation. However there is a conspiracy
in this theory: acausal propagation combined with
nonlocal interactions yield a causal theory, as exemplified by
the S-matrix.

\subsection{An acausal interacting theory}

Suppose that instead of (\ref{ourfourfermint}) we had attempted
to define an interacting homeotic theory from a {\it local}
four-fermi interaction: 
\be
H_I(t) = \int d^3x \H_{I}(x) = \int d^3x \,
(\bar{\psi}\gamma^{\mu}\psi)(\bar{\psi}\gamma_{\mu}\psi)\; .
\label{badfourfermint}
\ee
Despite the fact that this theory has a simple local interaction,
it exhibits strange properties. This theory preserves the global
$U(1)$ invariance, as demonstrated by the fact that the local
``current'' $J_D^{\mu}(x)=\bar{\psi}\gamma^{\mu}\psi$ commutes with the
charge $Q$ defined in (\ref{ourchargeb}). However this local current
is not conserved, and does not obey the current algebra relations
(\ref{hcurrentalg}). We find instead
\be
\left[ J^{\mu}(x), J^{\nu}(y) \right]_{x^0=y^0} = 
-im\left[D(x-y)+D(y-x)\right]\left(\bar{\psi}(x)
\sigma^{\mu\nu}\psi(y) + \bar{\psi}(y)\sigma^{\mu\nu}\psi(x)
\right)\; .
\label{badcurrents}
\ee
As a result, the causality condition Eq.(\ref{causalitycond}) fails
to hold, and the S-matrix is not well-defined. Put another way,
if we compute tree-level $2\rightarrow 2$ scattering of these
homeotic fermions using naive Feynman rules, we find that the
amplitudes violate crossing symmetry. Thus if the S-matrix were
well-defined, we would be exhibiting a simple local interaction that
violates crossing symmetry, which would be quite remarkable.

\section{$CPT$ violation and neutrinos}

The left-handed neutrinos of the Standard Model are Dirac-Weyl
fermion; they could not be represented as homeotic fermions unless
we drastically altered the gauge interactions of the Standard Model
to somehow make them compatible with the nonlocality of the
free homeotic lagrangian.

Because neutrinos have mass, it is natural to suppose that the
left-handed neutrinos $\nu_L^i(x)$, $i=1$,2,3,
have bilinear couplings to right-handed
neutrinos $N_R^i(x)$. These right-handed neutrinos are Standard
Model singlets, and might not carry any conserved local charges.
This fact is usually used as motivation for assuming that the
$N_R^i$ neutrinos have Majorana masses; instead we will use it
as motivation for assuming that the $N_R$ neutrinos are homeotic.

Consider the simplest possible model, where the $N_R^i$ neutrinos
are the right-handed components of free homeotic fermions $N^i$.
We assume the bilinear coupling
\be
m\int d^4x\, \left(\bar{\nu}_L(x)N_R(x) + \bar{N}_R(x)\nu_L(x) \right) \; ,
\label{mixedmassterm}
\ee
where we are now suppressing flavor indices for simplicity.
This coupling is both local and Lorentz invariant.

We can assume
that $m$ is parametrically small compared to other contributions
to the neutrino mass matrix, in which case it is sensible to
treat (\ref{mixedmassterm}) perturbatively, \ie , to consider
\be
m\int d^3x\, \left(\bar{\nu}_L(x)N_R(x) + \bar{N}_R(x)\nu_L(x) \right)
\label{mixedmasstermb}
\ee 
as part of the interaction hamiltonian density $\H_I(x)$.

Remarkably, the hybrid Dirac-homeotic theory thus defined obeys
the causality condition Eq.(\ref{causalitycond}). 
This can be worked out explicitly, but it is nearly obvious from
the anticommutation relations (\ref{hcomchiral}) for homeotic
fermions. We see there that right-handed homeotic fermions
have canonical anticommutation relations with other right-handed
homeotic fermions. Since the interaction hamiltonian density
constains only right-handed homeotic fermion fields (and Dirac-Weyl
fermion fields), the causality condition is obeyed just as it
is for an ordinary Dirac mass term.
This simple hybrid theory thus has a well-defined unitary Lorentz
invariant S-matrix. The lagrangian is nonlocal, but the nonlocality
appears only in the kinetic terms of the homeotic fermions.

This theory also violates $CPT$. This is apparent from the fact that
the $CPT$ symmetry operator is of necessity different for Dirac
theory and homeotic theory. We can see the $CPT$ violation explicitly
by evaluating (\ref{mixedmasstermb}) on-shell. For simplicity, suppose
that (\ref{mixedmasstermb}) is a perturbation on a theory of a Dirac
neutrino ($\nu_L$, $\nu_R$) and a homeotic neutrino ($N_L$, $N_R$)
with equal masses. Then on-shell (\ref{mixedmasstermb}) becomes:
\be
\displaystyle{ \int {d^3p\over (2\pi)^3}\, {1\over 2E_{\p}}
\sum_s \left[ m\left(a_{\p D}^{s\dagger}a_{\p h}^s -
b_{\p h}^{s\dagger}b_{\p D}^s \right)
+E_{\p}\left( b_{\p D}^sa_{-\p h}^s\e^{-2iE_{\p}t}
+ a_{\p D}^{s\dagger}b_{-\p h}^{s\dagger}\e^{2iE_{\p}t} \right)
 + h.c. \right] }\, ,
\label{cptvham}
\ee
where $a_{\p D}^{s\dagger}$, $b_{\p D}^{s\dagger}$ are Dirac creation
operators, and $a_{\p h}^{s\dagger}$, $b_{\p h}^{s\dagger}$ are
homeotic creation operators. Since $CPT$ exchanges
$a_{\p D}^{s\dagger}\leftrightarrow b_{\p D}^{s\dagger}$ and
exchanges $a_{\p h}^{s\dagger}\leftrightarrow b_{\p h}^{s\dagger}$,
the expression (\ref{cptvham}) is $CPT$ odd. 

Applied to neutrinos, this model produces precisely the $CPT$
violating neutrino mass spectra postulated in our
phenomenological work. Here we have obtained it from a simple
unitary causal framework.

\section{Discussion}
 
In this paper we have introduced what is certainly the simplest
Lorentz invariant field theory model for $CPT$ violating neutrino
oscillations. This model can be used to study a number of important
issues, such as $CPT$ violating baryogenesis, and off-shell neutrino
propagation. 

Homeotic fermions are interesting in their own right. They are in
some sense a unique alternative to Dirac fermions. They allow
interacting field theories which are nonlocal but causal.
There appears to be a deep connection -- a kind of duality --
between nonlocality in homeotic theory and ``negative frequency
spinors'' in Dirac theory. Homeotic supersymmetry is likely to
have novel properties, which could be of phenomenological importance.

\subsection*{Acknowledgements}
\noindent
We are grateful to Bill Bardeen, Liubo Borissov, Quico Botella,
Wally Greenberg, Howard Haber, Rob Myers and Arkady Vainshtein
for comments and suggestions. We thank our classical
consultant Maria Spiropulu for suggesting the term
``homeotic''.  We also thank the Aspen Center for Physics 
for providing a stimulating research enviroment.
This research was supported by the U.S.~Department of Energy
Grant DE-AC02-76CHO3000.

\end{document}